\documentclass[11pt, oneside]{article}

\usepackage{amsbsy,amssymb,amsmath,amsfonts,amsthm,latexsym,multicol,multirow,epsfig,color,subfigure,setspace,array,url,mathrsfs,bm,algpseudocode,algorithm}
\usepackage[stable]{footmisc}
\usepackage[authoryear]{natbib}
\usepackage{graphicx}
\usepackage{enumitem}
\usepackage{authblk}
\usepackage{xcolor}
\usepackage{verbatim} 
\usepackage{soul}
\usepackage{setspace}

\newtheorem{theorem}{Theorem}[section]
\newtheorem{lemma}{Lemma}[section]

\theoremstyle{definition}

\newcommand{\Y}{\bm{y}}

\def\T{{ \mathrm{\scriptscriptstyle T} }}

\def\H{{\mathcal{H}}}
\def\N{{\mathcal{N}}}

\def\T{{\intercal}} 
\def\RR{{\mathbb{R}}}
\def\Y{{\mathbf{Y}}}

\begin{document}
\title{Smoothing splines approximation using Hilbert curve basis selection}
\date{}

\author[1]{Cheng Meng}
\author[2]{Jun Yu}
\author[3]{Yongkai Chen}
\author[3]{Wenxuan Zhong}
\author[3,*]{Ping Ma}

\affil[1]{Center for Applied Statistics, Institute of Statistics and Big Data, Renmin University of China}
\affil[2]{School of Mathematics and Statistics, Beijing Institute of Technology}
\affil[3]{Department of Statistics, University of Georgia}
\affil[*]{Corresponding author: Ping Ma, pingma@uga.edu}

\maketitle

\doublespacing

\begin{abstract}
Smoothing splines have been used pervasively in nonparametric regressions. However, the computational burden of smoothing splines is significant when the sample size $n$ is large.
When the number of predictors $d\geq2$, the computational cost for smoothing splines is at the order of $O(n^3)$ using the standard approach.
Many methods have been developed to approximate smoothing spline estimators by using $q$ basis functions instead of $n$ ones, resulting in a computational cost of the order $O(nq^2)$.
These methods are called the basis selection methods. 
Despite algorithmic benefits, most of the basis selection methods require the assumption that the sample is uniformly-distributed on a hyper-cube. 
These methods may have deteriorating performance when such an assumption is not met.
To overcome the obstacle, we develop an efficient algorithm that is adaptive to the unknown probability density function of the predictors.
Theoretically, we show the proposed estimator has the same convergence rate as the full-basis estimator when $q$ is roughly at the order of $O[n^{2d/\{(pr+1)(d+2)\}}]$, where $p\in[1,2]$ and $r\approx 4$ are some constants depend on the type of the spline.
Numerical studies on various synthetic datasets demonstrate the superior performance of the proposed estimator in comparison with mainstream competitors.
\end{abstract}

\noindent%
\vfill

\newpage
\section{Introduction}
Smoothing spline estimators have been used pervasively in nonparametric regression models 
\begin{equation}\label{model1}
y_i=\eta(\bm{x}_i)+\epsilon_i \quad i=1,\ldots,n,
\end{equation}
where $y_i\in \mathbb{R}$ is the response, $\bm{x}_i \in\mathbb{R}^d$ is the predictor, $\eta$ is the unknown function to be estimated, and $\{\epsilon_i\}_{i=1}^n$ are i.i.d. normal random errors with zero mean and unknown variance $\sigma^2$ \citep{wahba1990spline,wang2011smoothing,gu2013smoothing,zhang2018smoothing}.
Despite their impressive performance,  smoothing splines suffer from a huge computational burden when the sample size $n$ is large.
Although univariate smoothing splines can be computed in $O(n)$ time \citep{reinsch1967smoothing}, in general cases when the number of predictors $d\geq2$, the classical method for calculating smoothing splines requires computing the inverse of a $n\times n$ matrix.
The standard algorithm for calculating matrix inversion requires $O(n^3)$ computational time.
To reduce such a huge computational cost, existing methods approximate smoothing spline estimators by using $q \ll n$ basis functions instead of $n$ ones. 
These methods are called the basis selection methods, which can reduce the computational cost to $O(nq^2)$.
Notice that one can further refine the order $O(n^3)$ and $O(nq^2)$ to $o(n^3)$ and $o(nq^2)$, respectively, using Strassen algorithm, Coppersmith–Winograd algorithm or Optimized CW-like algorithms \citep{bernstein2009matrix,golub2013matrix}.
These algorithms are beyond the scope of this paper.

Various basis selection methods have been proposed.
\citet{luo1997hybrid} and \citet{zhang2004variable} selected the basis functions through variable selection techniques.
\citet{hastie1996pseudosplines} and \citet{ruppert2002selecting} considered pseudosplines, also called P-splines, which utilize $q$ fixed basis functions to approximate splines.
Such fixed basis functions are also called knots and differ from the construction of the basis functions in smoothing splines.
\citet{he2001data,sklar2013nonparametric}, and \citet{yuan2013adaptive} considered the cases that the regression function has non-homogeneous smoothness across the design space.
They developed data-driven methods to select basis functions or knots, such that the selected ones are adaptive to non-homogeneous smoothness of the regression function.

There also exist other strategies that aim to approximate splines or other nonparametric regression estimators in a computationally efficient manner through parallel computing.
\cite{zhang2013divide} and \cite{zhang2015divide} studied the divide-and-conquer kernel ridge regression (dacKRR), and showed that it achieves minimax optimal convergence rates under relatively mild conditions.
\cite{wood2017generalized} accelerated the fitting of penalized regression spline based generalized additive models.
They showed that their method could run reliably and efficiently on a desktop workstation for $d$ up to $10^4$ and $n$ up to $10^8$.
\cite{xu2018divide} and \cite{xu2019distributed} considered the problem of how to estimate the tuning parameter effectively for dacKRR.
They proposed a variant of the generalized cross-validation for dacKRR, and showed that their proposed technique is computationally scalable for massive datasets and is asymptotically optimal under mild conditions.
\cite{shang2017computational} analyzed the theoretical properties of one-dimensional smoothing splines under the divide-and-conquer setting.
\cite{liu2018many} and \cite{liu2020nonparametric} studied the theoretical properties of dacKRR respecting the number of machines.
They showed that there exists a specific bound for the number of machines in order to let the dacKRR estimators to achieve statistical minimax.
\citep{shang2019nonparametric} developed scalable Bayesian inference procedures for a general class of nonparametric regression models using distributed learning. 
In practice, it is possible to combine the aforementioned parallel-based strategies with the proposed method for more computational savings.

One fundamental question for basis selection methods is how to determine the size of $q$, which balances the trade-off between the computation time and the prediction accuracy.
In this paper, we focus on the widely-used asymptotic criterion, which aims to determine the smallest order of $q$ such that the $q$-basis estimator converges to the true function $\eta$ at the same rate as the full-basis estimator.
\cite{zhou2001spatially} proposed an estimator for regression spline using the spatial adaptive basis functions. 
This method has been applied in univariate cases; however, it is not clear whether it can be extended to multivariate cases.
\cite{xiao2013fast} proposed an estimator for P-spline under the scenario that the observations are supported on a $n_1\times n_2$ grids, and showed that the essential number of basis $q=n_1n_2/4$.
One limitation of their estimator is that it can only be applied in the cases when the observations are supported on a two-dimensional grid.
\cite{gu2002penalized} and \cite{ma2015efficient} developed the uniform basis selection method and the adaptive basis selection method, respectively.
Both methods require $q$ roughly be of the order $O\{n^{2/(pr+1)}\}$, where $p\in[1,2]$ and $r\approx 4$ are some constants depend on the type of the spline.
We provide a discussion on these two constants in Section~4.
Recently, \cite{meng2020more} proposed a more efficient basis selection method that only require $q$ roughly be of the order $O\{n^{1/(pr+1)}\}$, when $d\leq pr+1$.
Their method aims to select approximately uniformly-distributed observations by utilizing space-filling designs or low-discrepancy sequences, resulting in a faster convergence rate compared with the uniform basis selection method.

\begin{figure}[!ht]\centering
        \includegraphics[scale = .8]{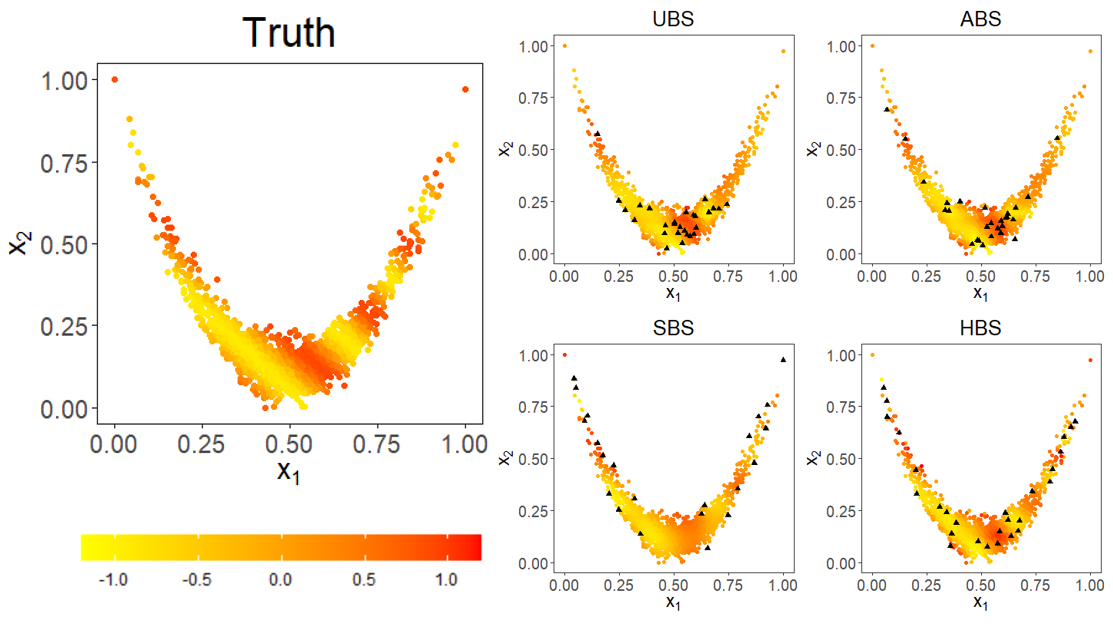} 
        \caption{The leftmost panel shows the heat map for the true function. The heat maps for the spline estimates based on the uniform basis selection method (UBS), the adaptive basis selection method (ABS), the space-filling basis selection method (SBS), and the proposed Hilbert basis selection method (HBS) are presented in the other panels, respectively. Black triangles are the selected basis functions. We observe that the proposed method outperforms the other methods in approximating the true function.} \label{fig3}
\end{figure}

Despite algorithmic benefits, the key to the success of most of the existing basis selection methods depends on the assumption that the sample is uniformly-distributed in a hyper-cube or a hyper-rectangular.
In practice, most basis selection methods may suffer from deteriorating performance when such an assumption is not met.
We now demonstrate the case that the sample is not uniformly-distributed using a toy example.
In this example, we generate two thousand data points from a banana-shape distribution on $[0,1]^2$, and we show the heat map of the true response surface $y=\sin\{20(x_1+x_2)\}$ in the leftmost panel of Fig.~\ref{fig3}.
The marginal distribution of such banana-shape distribution conditional on $x_1$ is a Gaussian distribution; thus, more data points are located in the middle than on the boundary.
We compare the proposed method, denoted by HBS, with the mainstream competitors, which includes the uniform basis selection \citep{gu2002penalized}, the adaptive basis selection \citep{ma2015efficient}, and the space-filling basis selection \citep{meng2020more}.
We set $q=5\times(2000)^{2/9}\approx27$ for all basis selection methods, and we mark the selected basis functions as black triangles. 
The right four panels of Fig.~\ref{fig3} show the heat maps of the spline estimates of all four basis selection methods, respectively.
We observe the uniform basis selection method and the adaptive basis selection method perform similarly: both select very few basis functions on the boundary. 
As a result, these two methods fail to capture the periodic pattern of the response surface on the boundary.
In contrast, the space-filling basis selection method selects very few basis functions in the middle, resulting in degenerated performance in such an area.
These observations suggest that the performance of a basis selection method may deteriorate significantly when the sample is not uniformly generated in a hypercube.
Finally, we observe the proposed method is adaptive to the arbitrary distribution of the sample, resulting in the best estimation of the true function, compared with other methods.

In practice, the distribution of the sample is almost always unknown to practitioners. 
The basis selection method hence is highly desirable to be robust to arbitrary distribution of the sample. 
To achieve the goal, it is suggested in Chapter~4 of \citet{gu2013smoothing} to select the basis functions corresponding to roughly equally-spaced observations, even when the sample is not uniformly-distributed.
Analogously, \cite{eilers2010splines} found that equally-spaced knots, which can be regarded as the basis functions here, are always preferred in practice.
These discoveries are consistent with the common key idea in importance sampling techniques, which are widely-used for variance-reduction in numerical integration \citep{liu1996metropolized,liu2008monte}.
We now briefly introduce such an idea in the following.

Let $f$ be an integrand and $g$ be a probability density function on $\Omega\subseteq\mathbb{R}^d$.
To estimate the integration $\int_\Omega f(\bm{x})g(\bm{x})\mbox{d}\bm{x}$, one can simply generate an i.i.d. sample $\{\bm{x}_i\}_{i=1}^n$ from $g$, then calculate the mean of $\{f(\bm{x}_i)\}_{i=1}^n$. 
Instead, one can also generate an i.i.d. sample from a probability density function $h$, then calculate the mean of $\{f(\bm{x}_i)g(\bm{x}_i)/h(\bm{x}_i)\}_{i=1}^n$.
\cite{kahn1953methods} showed when both $f$ and $g$ are known, the optimal $h(\bm{x})$ in terms of variance-reduction is proportional to $|f(\bm{x})|g(\bm{x})$, outside of trivial cases where $\int |f(\bm{x})|g(\bm{x})\mbox{d}\bm{x}=0$.
The intuition is that $h(\bm{x})$ needs to have sufficiently large value for the $x$ such that $|f(\bm{x})|g(\bm{x})$ is close to zero.
Otherwise, if $h(\bm{x})$ is extremely small for such $\bm{x}$, the variance of $\mathbb{E}_g(|f(\bm{x})|)=|f(\bm{x})|g(\bm{x})/h(\bm{x})$ can be inflated to be arbitrarily large.
Consequently, in the cases when either $f$ or $g$ is unknown, a safe strategy is to let $h$ be the uniform distribution on $\Omega$; thus simply avoids the scenario that $h(\bm{x})$ is extremely small for any $\bm{x}\in \Omega$.

Inspired by such a strategy in importance sampling techniques, we propose a novel basis selection method by selecting the basis functions corresponding to roughly equally-spaced observations.
To achieve the goal, we develop an efficient algorithm that utilizes the Hilbert space-filling curve.
The proposed algorithm can be used to select a uniformly-distributed subsample without knowing the probability density function of the predictors.
Theoretically, we show the proposed estimator has the same convergence rate as the full-basis estimator.
Furthermore, we show the order of $q$ for the proposed method is reduced from roughly $O(n^{2/(pr+1)})$ in the uniform basis selection method to roughly $O(n^{2d/\{(pr+1)(d+2)\}})$.
To the best of our knowledge, in the cases when the sample follows an arbitrary distribution, the proposed estimator is the one that requires the smallest order of $q$.
Numerical studies on various synthetic datasets demonstrate the superior performance of the proposed estimator in comparison with mainstream competitors.

Although we mainly focus on smoothing splines in this paper, it is possible that the proposed method could also accelerate the estimation of other nonparametric regression estimators, includes the thin plate regression splines, kernel ridge regression and etc \citep{geer2000empirical,wood2003thin,gyorfi2006distribution,wasserman2006all,hastie2009elements,yang2017randomized}.
Some simulation results are provided in Supplementary Material to support this claim.

The rest of the paper is organized as follows. We review smoothing splines and basis selection methods in Section~\ref{section:2}. In Section~\ref{section:3}, we first introduce the Hilbert curve and some of its properties. We then introduce our basis selection method utilizing the Hilbert curve. In Section~\ref{section:4}, we present theoretical properties of the proposed method. We evaluate the empirical performance of the proposed method via extensive simulation studies and a real-world data analysis in Section~\ref{section:5} and Section~\ref{section:6}, respectively.
Section~7 includes some discussion of the paper.
Proofs of the theorems are collected in Supplementary Material.

\section{Preliminaries}
\label{section:2}

\subsection{Background of Smoothing Splines}
To estimate the unknown function $\eta$ in Model~(\ref{model1}), a common strategy is to minimize the penalized least squares criterion \citep{wahba1990spline, wang2011asymptotics,gu2013smoothing, wang2013smoothing},  
\begin{eqnarray}\label{eqn1}
\frac{1}{n}\sum_{i=1}^n\{y_i-\eta(\bm{x}_i)\}^2+\lambda J(\eta),
\end{eqnarray}
where $J(\cdot)$ is a squared semi-norm, and $J(\eta)$ is called the roughness penalty.
The $\lambda$ here is the smoothing parameter, which balances the trade-off between the goodness-of-fit of the model and the roughness of the function $\eta$. 
Such $\lambda$ can be selected based on the generalized cross-validation (GCV) criterion \citep{wahba1978smoothing}.
\cite{xu2018divide} and \cite{xu2019distributed} generalized the GCV criterion to the setting of distributed learning.
Recently, \cite{sun2021asymptotic} proposed a more efficient approach for accelerating the calculation of $\lambda$.

In this paper, we focus on minimizing the objective function~(\ref{eqn1}) in a reproducing kernel Hilbert space, resulting in a smoothing spline estimate for $\eta$.
Let $\H=\{\eta:J(\eta)<\infty\}$ be a reproducing kernel Hilbert space and $\N_J=\{\eta:J(\eta)=0\}$ be the null space of $J(\eta)$.
Let $\N_J$ be a $m$-dimensional linear subspace of $\H$ and $\{\xi_i\}_{i=1}^m$ be a set of basis for $\N_J$.
Moreover, let $\H_J$ to denote the orthogonal complement of $\N_J$ in $\H$ such that $\H=\N_J\oplus \H_J$.  
It can be shown that $\H_J$ is a still a reproducing kernel Hilbert space, and we use $R_J(\cdot,\cdot)$ to denote the reproducing kernel of $\H_J$.
Let $\Y=(y_1,\ldots,y_n)^\T$ be the response vector, $\textbf{S}\in\RR^{n\times m}$ be a matrix where the $(i,j)$-th element equals $\xi_j(\bm{x}_i)$, and $\textbf{R}\in\RR^{n\times n}$ be a matrix where the $(i,j)$-th element equals $R_J(\bm{x}_i,\bm{x}_j)$.
According to the representer theorem \citep{wahba1990spline}, the minimizer of the objective function (\ref{eqn1}) in the space $\H$ takes the form $\eta(\bm{x})=\sum_{k=1}^m \alpha_k\xi_k(\bm{x})+\sum_{i=1}^n \beta_i R_J(\bm{x}_i,\bm{x}).$
Let $\bm{\alpha}=(\alpha_1,\ldots,\alpha_m)^\T$ and $\bm{\beta}=(\beta_1,\ldots,\beta_n)^\T$ be the coefficient vectors.
With trivial modification, it can be shown that finding the minimizer of the objective function~(\ref{eqn1}) is equivalent to solving
\begin{equation}\label{eqn4}
(\hat{\bm{\alpha}}, \hat{\bm{\beta}})=
\underset{\bm{\alpha}\in\mathbb{R}^m,\mbox{ } \bm{\beta}\in\mathbb{R}^n}{\mathrm{argmin}}\frac{1}{n}(\Y-\textbf{S}\bm{\alpha}-\textbf{R}\bm{\beta})^\T(\Y-\textbf{S}\bm{\alpha}-\textbf{R}\bm{\beta})+\lambda \bm{\beta}^\T \textbf{R}\bm{\beta}.
\end{equation} 

Although the solution of the minimization problem~(\ref{eqn4}) has a closed form \citep{gu2002penalized,ma2015efficient}, the computational cost for calculating the solution is of the order $O(n^3)$, in a general case where $n\gg m$ and $d\geq2$.

\subsection{Basis Selection Methods}
To alleviate the computation burden for smoothing splines, various basis selection methods have been developed. 
These methods are of similar nature to the subsampling methods, which are widely used in large-scale data analysis \citep{mahoney2011randomized,drineas2012fast,ma2015statistical,ma2015leveraging,meng2017effective,zhang2018statistical,ai2019optimal,xie2019online,ma2020asymptotic,yu2020optimal,AI2020101512,meng2020lowcon,zhong2021model}.
We refer to \cite{li2020modern} for a recent review.

The standard basis selection method works as follows.
One first use subsampling techniques to select a subsample $\{\bm{x}_i^*\}_{i=1}^q$ from the observed sample $\{\bm{x}_i\}_{i=1}^n$.
The selected subsample is then used to construct the so-called effective model space $\H_E=\N_J\oplus \mbox{span}\{R_J(\bm{x}_i^*,\cdot),i=1,\ldots,q\}$.
Finally, the objective function~(\ref{eqn1}) is minimized in the effective model space $\H_E$, and the solution thus can be written as
$\eta_E(\bm{x})=\sum_{k=1}^m \alpha_k\xi_k(\bm{x})+\sum_{i=1}^q \beta_i R_J(\bm{x}^*_i,\bm{x})$.
Analogous to Equation (\ref{eqn4}), the coefficients $\bm{\alpha}_E=(\alpha_1,\ldots,\alpha_m)^\T$ and $\bm{\beta}_E=(\beta_1,\ldots,\beta_q)^\T$ can be obtained through solving
\begin{eqnarray}\label{eqn5}
(\hat{\bm{\alpha}}_E, \hat{\bm{\beta}}_E)=
\underset{\bm{\alpha}_E\in\RR^m,\mbox{ }\bm{\beta}_E\in\RR^q}{\mathrm{argmin}}\frac{1}{n}(\Y-\textbf{S}\bm{\alpha}_E-\textbf{R}_*\bm{\beta}_E)^\T(\Y-\textbf{S}\bm{\alpha}_E-\textbf{R}_*\bm{\beta}_E)+\lambda \bm{\beta}_E^\T \textbf{R}_{**}\bm{\beta}_E,
\end{eqnarray} 
where $\textbf{R}_*\in \mathbb{R}^{n\times q}$ is a matrix where the $(i,j)$-th element equals $R_J(x_i,x_j^*)$ and $\textbf{R}_{**} \in \mathbb{R}^{q\times q}$ is a matrix where the $(i,j)$-th element equals $R_J(x_i^*,x_j^*)$. 
In general cases where $m\ll q\ll n$, solving the optimization problem~(\ref{eqn5}) requiring only $O(nq^2)$ computation time, which is a significant reduction compared with $O(n^3)$.

Despite algorithmic benefits, most of the existing basis selection methods heavily rely on the condition that the sample is uniformly-distributed on a hyper-cube.
When such a condition is not met, they may suffer from deteriorating performance, as shown in Fig~\ref{fig3}.
Recall that a common strategy in importance sampling techniques is to select a roughly uniformly-distributed subsample, which tends to be beneficial for numerical integration.
Such a strategy motivates us to select the basis functions corresponding to roughly equally-spaced observations, even when the sample is not uniformly-distributed.
Intuitively, such a goal can be easily achieved when $d=1$, in which cases one can first divide the sample space into equally-spaced bins, and then select an equal number of observations within each bin. 
The selected subsample is roughly uniformly-distributed when the number of bins is carefully determined.
Unfortunately, such a naive strategy is not easily extendable to the cases that $d\geq 2$, due to the curse-of-dimensionality.

To overcome the barrier, a natural strategy is to find a continuous mapping $F:\Omega\rightarrow\RR$ that preserves local structures, where $\Omega\subset\RR^d$ is a bounded design space.
In other words, we aim to find a mapping $F$ such that, for any $\bm{x}_i,\bm{x}_j\in\Omega$, $i,j\in\{1,\ldots,n\}$, a small value of $||\bm{x}_i-\bm{x}_j||$ is associated with a small value of $||F(\bm{x}_i)-F(\bm{x}_j)||$, where $||\cdot||$ represents the Euclidean norm.
Loosely speaking, let $\{F(\bm{x}_i^*)\}_{i=1}^q$ be a roughly uniformly-distributed subset selected from $\{F(\bm{x}_i)\}_{i=1}^n$, the subsample $\{\bm{x}_i^*\}_{i=1}^q$ thus tends to be uniformly-distributed in $\Omega$.
One family of the mappings that approximately achieve this goal is the family of space-filling curves, which include the Hilbert curve, the Peano curve, and the Z-order curve \citep{sagan2012space}.
We develop a novel basis selection method utilizing space-filling curves, as detailed in the next section.

\section{Basis Selection using Space-Filling Curves}
\label{section:3}
\subsection{Hilbert Curves}

Space-filling curves have long been studied in mathematics and have become important computational tools since the 1980s \citep{bader2012space,li2020stratification}.
Nowadays, space-filling curves have been widely used for computer graphics, approximately nearest neighbor searching, solving partial differential equations, and so on \citep{zumbusch2012parallel}.
We now briefly introduce the Hilbert curve, a representative of space-filling curves, and some of its properties that we need.
The formal definition of the Hilbert curve is relegated to Supplementary Material.
Other space-filling curves enjoy similar properties, and we refer to \citet{sagan2012space,zumbusch2012parallel} for more details.

\begin{figure}[ht]
\begin{center}
        \includegraphics[scale = 0.45]{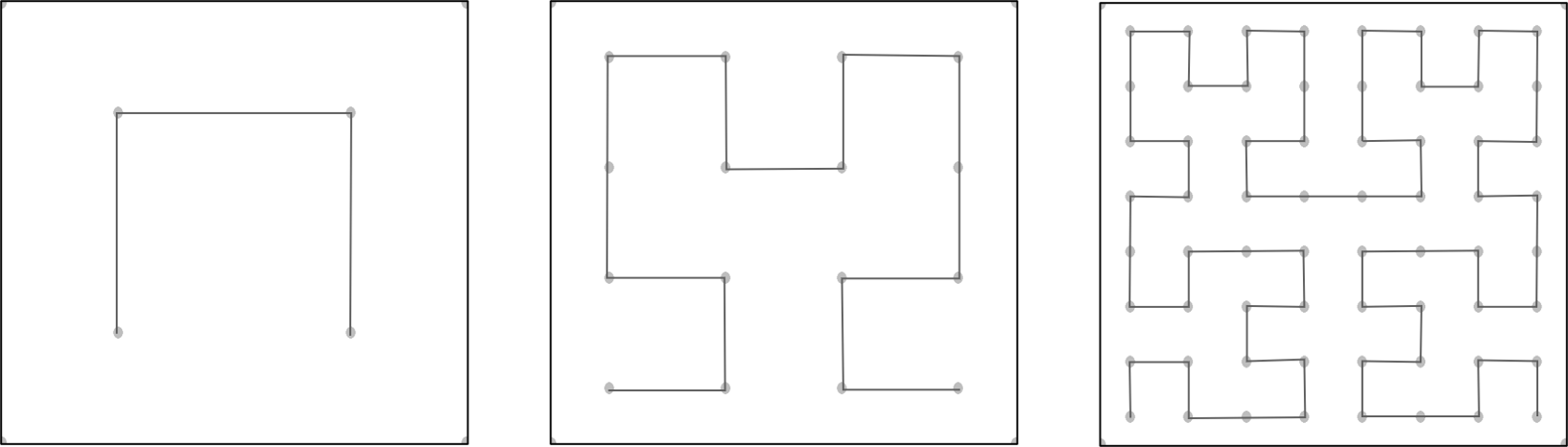}
        \caption{Illustration for the first three stages of the Hilbert’s space filling curves when $d=2$. From left to right, each panel shows $H_1$, $H_2$, and $H_3$, respectively.} \label{fig1}
\end{center}
\end{figure}

We first introduce a sequence of the so-called Hilbert space-filling curves, denoted by $\{H_k\}_{k=1}^\infty$.
Intuitively, for each $k$, the $k$-th Hilbert space-filling curve $H_k$ is a bijection between a partition of $[0,1]$ and a partition of $[0,1]^d$.
In particular, the curve $H_k$ partitions both $[0,1]$ and $[0,1]^d$ into $(2^k)^d$ blocks, respectively, and construct a bijection between these blocks.
Figure~\ref{fig1} illustrates how a partition of $[0,1]$ is mapped to a partition of $[0,1]^2$ using $H_1$, $H_2$, and $H_3$, respectively.
The Hilbert curve is defined as $H(x)=\lim_{k\rightarrow\infty}H_k(x)$, which becomes a mapping from $[0,1]$ to $[0,1]^d$.
It is well-known that the Hilbert curve $H$ enjoys the locality-preserving property \citep{zumbusch2012parallel}.
In particular, for any $x,y\in[0,1]$, one has
\begin{equation}\label{hcurve}
||H(x)-H(y)||\leq2\sqrt{d+3}|x-y|^{1/d}. 
\end{equation}

Inequality~(\ref{hcurve}) indicates a small value of $|x-y|$ is associated with a small value of $||H(x)-H(y)||$, despite the fact that the converse can't always be true.
Inequality~(\ref{hcurve}) inspired us to select approximately equally-spaced observations using the Hilbert curve $H$.
In practice, $H_k$ is used as a surrogate of $H$ due to the computational concern.

\subsection{Hilbert Basis Selection Method}
We develop a novel basis selection method utilizing Hilbert space-filling curves, called the Hilbert basis selection method.
The proposed method works as follows.
We first scale the sample $\{\bm{x}_i\}_{i=1}^n\in\mathbb{R}^d$ to $[0,1]^d$ as a pre-processing step.  
Recall that the Hilbert space-filling curve $H_k$ partitions both $[0,1]$ and $[0,1]^d$ into $2^{kd}$ blocks, denoted by $\{c'_j\}_{j=1}^{2^{dk}}$ and $\{c_j\}_{j=1}^{2^{dk}}$, respectively, and construct a bijection between these blocks.
For any data point $\bm{x}\in[0,1]^d$, we assign $\bm{x}$ to its corresponding block $c_j$ in $[0,1]^d$, $j\in\{1,\ldots,2^{kd}\}$, then map $\bm{x}$ to the center of the block $c_j'=H_k^{-1}(c_j)$.
This is to say, all the data points that belong to the same block are mapped to the same point in $[0,1]$.
Next, given a positive integer $C$, we draw the histogram for the mapped data points with $C$ bins. 
Let $\tilde{C}$ be the number of non-empty bins and $q$ be the subsample size.
We then randomly select roughly $[q/\tilde{C}]$ data points from each non-empty bin. 
The subsample corresponding to the selected data points is used to construct the effective subspace $\H_E$.
Finally, we calculate the smoothing spline estimator $\eta_E$ in such a subspace. 
The algorithm is summarized below.

\begin{algorithm}
\caption{Hilbert basis selection method}
\begin{tabbing}
   \qquad \enspace \textit{Step 1.} The sample $\{\bm{x}_i\}_{i=1}
^n$ is first scaled to $[0,1]^d$.\\
   \qquad \enspace \textit{Step 2.} Calculate the bijection between $\{c'_j\}_{j=1}^{2^{dk}}$ and $\{c_j\}_{j=1}^{2^{dk}}$ using the Hilbert space-filling\\
   \qquad \qquad curve $H_k$.\\
   \qquad \enspace \textit{Step 3.} For each data point $\bm{x}_i$, $i=1,\ldots,n$, suppose $\bm{x}_i$ belongs to the block $c_j$,\\
   \qquad \qquad map $\bm{x}_i$ to the center of the block $c_j'=H_k^{-1}(c_j)$.\\
   \qquad \enspace \textit{Step 4.} Draw a histogram for the mapped points with $C$ bins; \\
   \qquad \qquad let $\tilde{C}$ be the number of non-empty bins.\\
   \qquad \enspace \textit{Step 5.} Randomly select roughly $[q/\tilde{C}]$ number of data points from each non-empty bin;\\
   \qquad \qquad  let $\{\bm{x}_i^*\}_{i=1}^q$ to denote the selected ones. \\
   \qquad \enspace \textit{Step 6.} Minimize the objective function (\ref{eqn1}) over the effective subspace \\
   \qquad \qquad $\H_E=\mathcal{N}_J \oplus \mbox{span}\{R_J(\bm{x}_i^*,\cdot), i=1,\ldots,q\}.$
\end{tabbing}
\end{algorithm}

\begin{figure}[ht]\centering
        \includegraphics[scale = 0.45]{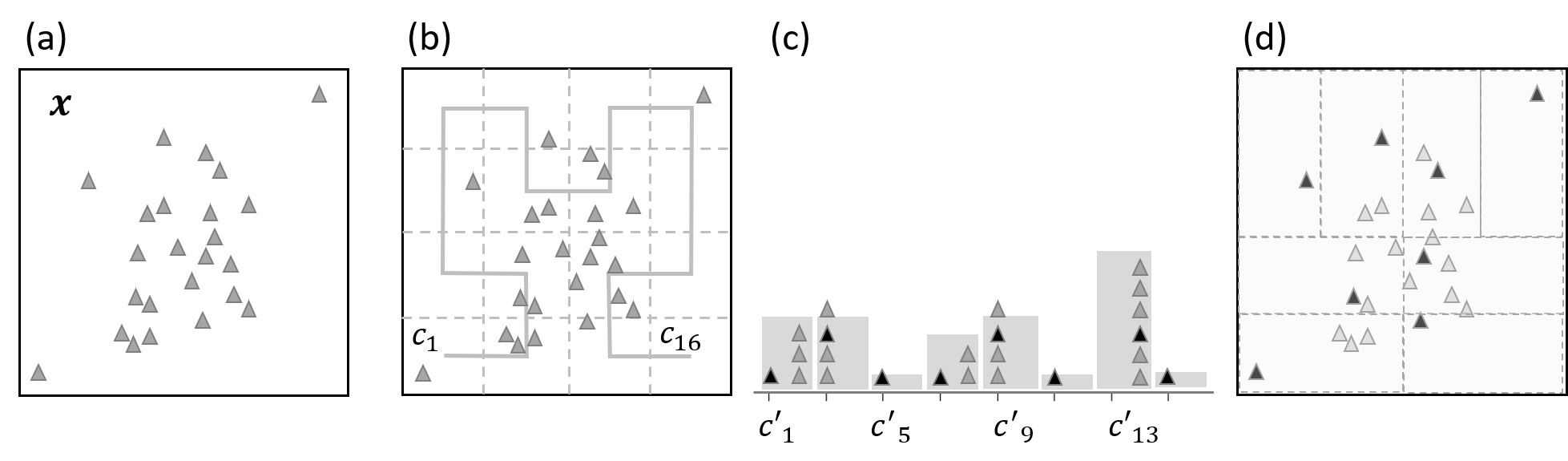}
        \caption{Illustration for Algorithm~1.} \label{fig2}
\end{figure}

Figure~\ref{fig2} gives an illustration of Algorithm~1, in which all data points are shown in  Fig.~\ref{fig2}(a).
We set $k=d=2$ in Algorithm~1, resulting in
$2^{2\times2}=16$ blocks in Fig.~\ref{fig2}(b), denoted by $c_1,\ldots,c_{16}$, respectively.
All the data points are then mapped to the center of $c_j'$s, $j=1,\ldots,16$, as shown in Fig.~\ref{fig2}(c).
Let the subsample size $q=8$.
We then draw $C=8$ bins for the histogram in Fig.~\ref{fig2}(c), resulting in $\tilde{C}=8$ non-empty bins.
We then randomly select $\tilde{C}/q=1$ data point from each bin.
The selected data points are labeled as black triangles in both Fig.~\ref{fig2}(c) and Fig.~\ref{fig2}(d).
Note that each bin in Fig.~\ref{fig2}(c) is associated with a ``meta-block", as illustrated in Fig.~\ref{fig2}(d).
As a result, when $\tilde{C}=q$, Algorithm~1 ensures none of the two selected data points lie in the same meta-block, and thus the selected subsample tend to be equally-spaced.

The choice of $k$ is a key to Algorithm~1, while the performance of Algorithm~1 is not sensitive to the choice of $k$, as long as $k$ is not too small.
The reasons are as follows.
Recall that for dimension $d$, the curve $H_{k_0}$ with respect to (w.r.t.) a positive integer $k_0$ partitions the interval $[0,1]$ into $2^{dk_0}$ blocks, denoted by $\{c'_j\}_{j=1}^{2^{dk_0}}$.
One nice property of the Hilbert space-filling curve is that, suppose one data point $\bm{x}$ is mapped into a block using the curve $H_{k_0}$, then for any $k\geq k_0$, the curve $H_{k}$ always map $\bm{x}$ into the same block, despite the fact that its position within the block may vary.
Note that Algorithm~1 draws a histogram for the mapped data points with $C$ bins and randomly selects several data points from each of the non-empty bins.
As a result, when $C$ is properly chosen and is fixed, the value of $k\geq k_0$ does not affect such a histogram, and thus does not affect the result of Algorithm~1.

The computational cost for Algorithm~1 mainly resides in Step 2 and Step 6.
It can be shown that the computational cost for Step 2 is of the order $O(n)$, which is negligible compared to the computational cost for Step 5, which is of the order $O(nq^2)$.
In sum, analogous to other basis selection methods, the overall computational cost for Algorithm~1 is at the order of $O(nq^2)$.


\section{Convergence Rates for Function Estimation}
\label{section:4}
Let $f_X$ be the probability density function of the predictors defined on $\Omega$.
We require $\Omega$ to be bounded, and without loss of generality, we assume $\Omega\subseteq[0,1]^d$.
Let $V(g)=\int_{\Omega} g^2f_X\mbox{d}\bm{x}$. 
A function $f(x)$ defined on $\Omega$ is said to be Lipschitz continuous, if for any $\bm{x},\bm{y}\in\Omega$, there exist a constant $M$ such that $|f(\bm{x})-f(\bm{y})|\leq M||\bm{x}-\bm{y}||$, where $||\cdot||$ is the Euclidean norm.
We introduce some essential regularity conditions in the following.

\begin{itemize}[noitemsep]
    \item Condition 1. The function $V$ is completely continuous with respect to $J$; 
    \item Condition 2. For some $\beta>0$ and $r>1$, $\rho_\nu>\beta\nu^r$ for sufficiently large $\nu$;
    \item Condition 3. For all $\mu$ and $\nu$, $\mbox{var}\{\phi_\nu(\bm{x})\phi_\mu(\bm{x})\}\le B$, where $\phi_\nu$, $\phi_\mu$ are the eigenfunctions associated with $V$ and $J$ in $\H$, $B$ denotes a positive constant.
    \item Condition 4. For all $\mu$ and $\nu$, $\phi_\mu(\bm{x})\phi_\nu(\bm{x})\in L^2(\Omega)$, and is Lipschitz continuous. 
    \item Condition 5. Assume that $\max\{ (qn_i)/n\}_{i=1}^C=O_p(1),$ where $n_i$ is the number of observations 
    in the $i$th bin.
    \item Condition 6. As $n\rightarrow\infty$, $q^{1+2/d}=O(n)$.
\end{itemize}

Condition 1 implies that there exist a sequence of eigenfunctions $\phi_\nu\in\H$ and the associated sequence of eigenvalues $\rho_\nu \uparrow \infty$ satisfying $V(\phi_\nu,\phi_\mu)=\delta_{\nu\mu}$ and $J(\phi_\nu,\phi_\mu)=\rho_\nu\delta_{\nu\mu}$, where $\delta_{\nu\mu}$ is the Kronecker delta.
The growth rate of $\rho_\nu$ is closely related to the convergence rate of smoothing spline estimates~\citep{gu2013smoothing}. 
Condition 1 can be verified under some special cases when the eigenfunctions are available in explicit forms. 
Consider $J(\eta)=\int_0^1 \eta^{''}\mbox{d}x$, where $\eta$ is a periodic function on $[0,1]$. 
The eigenfunctions $\phi_\nu$s are the sine and cosine functions in such a case, and thus Condition 1 holds naturally.
We refer to Section 9.1 of \citet{gu2013smoothing} for more details on the construction of the eigenfunctions.
In general, Conditions 1--3 are widely-used in the asymptotic analysis for smoothing spline estimates, and we refer to \citet{gu2013smoothing,ma2015efficient} for more technical discussion of these conditions.
Condition 4 is satisfied naturally for various choices of eigenfunctions. 
Condition 5 naturally holds for the regular sampling in $[0,1]^d.$ 
Moreover, Condition 5 prevents some extreme cases of the probability density function $f_X$.
For example, when the one-dimensional data points $\{H_k^{-1}(x_i)\}_{i=1}^n$ follow the Dirac delta function, one has $\max\{ (qn_i)/n\}_{i=1}^C=qn/n=q$, which is in conflict with Condition 5.
Finally, Condition 6 naturally holds when the number of basis $q$ is not too large.
For example, when $d\geq 2$, Condition~6 holds when $q=O(n^{1/2}).$

Recall that $\{\bm{x}_j^*\}_{j=1}^q$ is a subsample selected by the proposed algorithm.
Moreover, $C$ and $\tilde{C}$ are the number of bins and the number of non-empty bins in Step 4 of Algorithm~1, respectively.
For brevity, throughout this section, we assume only one data point is selected from each non-empty bin; that is, we assume $q=\tilde{C}$.
The extensions to more general cases where $q/\tilde{C}=O(1)$ are straightforward.
We let $\tilde{n}_j$ to denote the number of data points within the bin that $\bm{x}_j^*$ lies in.
Consider the estimator $\sum_{j=1}^q\{\tilde{n}_j/n\}\phi_\nu(\bm{x}_j^*)\phi_\mu(\bm{x}_j^*)$.
Intuitively, such an estimator can be regarded as the mean estimator of the stratified sampling. 
This is because, for $j=1,\ldots, q$, $\{\tilde{n}_j/n\}\phi_\nu(\bm{x}_j^*)\phi_\mu(\bm{x}_j^*)$ can be regarded as the sample mean of the $j$th strata.
The following lemma gives the convergence rate of the selected subsample in terms of numerical integration. 
All the proofs throughout this section are relegated to the Supplementary Material.

\begin{lemma}\label{lem1}
Under Conditions 4--6, for all $\mu$ and $\nu$, $\sum_{j=1}^q\{\tilde{n}_j/n\}\phi_\nu(\bm{x}_j^*)\phi_\mu(\bm{x}_j^*)$ is an asymptotically unbiased estimate for $\int_{\Omega}\phi_\nu(\bm{x})\phi_\mu(\bm{x}) f_X(\bm{x})\mbox{d}\bm{x}$.
Furthermore, when $\Omega\subseteq[0,1]^d$, we have
\begin{eqnarray*}
\left\{\int_{\Omega}\phi_\nu(\bm{x})\phi_\mu(\bm{x}) f_X(\bm{x}) \mbox{d}\bm{x}-\sum_{j=1}^q\{\tilde{n}_j/n\}\phi_\nu(\bm{x}_j^*)\phi_\mu(\bm{x}_j^*)\right\}^2=O_p(q^{-1-2/d}).
\end{eqnarray*}

\end{lemma}

Lemma \ref{lem1} shows the advantage of $\{\bm{x}_i^*\}_{i=1}^q$ over a randomly selected subsample $\{\bm{x}_i^+\}_{i=1}^q$. To be specific, for all $\mu$ and $\nu$, as a direct consequence of Condition 3, which assumes that the variance of $\phi_\nu(\bm{x})\phi_\mu(\bm{x})$ is finite, we have
$$\mathbb{E}\left[\int_{\Omega}\phi_\nu(\bm{x})\phi_\mu(\bm{x}) f_X(\bm{x})\mbox{d}\bm{x} -\frac{1}{q}\sum_{j=1}^q\phi_\nu(\bm{x}_j^+)\phi_\mu(\bm{x}_j^+)\right]^2=O(q^{-1}).$$ 
Consequently, Lemma \ref{lem1} suggests that one can approximate the integration $\int_{\Omega}\phi_\nu(\bm{x})\phi_\mu(\bm{x}) f_X(\bm{x}) \mbox{d}\bm{x}$ more effectively, by calculating $\sum_{j=1}^q(n_j/n)\phi_\nu(\bm{x}_j^*)\phi_\mu(\bm{x}_j^*)$ instead of $\sum_{j=1}^q\phi_\nu(\bm{x}_j^+)\phi_\mu(\bm{x}_j^+)/q$.
Lemma~1 paves the way for our main theorem below.

\begin{theorem}\label{thm1}
Suppose $\sum_i\rho_i^pV(\eta_0,\phi_i)^2<\infty$ for some $p\in[1,2]$.
Under Conditions 1--6, as $\lambda \rightarrow 0$ and $q^{1+2/d}\lambda^{2/r} \rightarrow \infty$, we have $(V+\lambda J)(\hat{\eta}_E-\eta_0)=O_p(n^{-1}\lambda^{-1/r}+\lambda^p)$.
In particular, if $\lambda \asymp n^{-r/(pr+1)}$, the estimator achieves the optimal convergence rate $$(V+\lambda J)(\hat{\eta}_E-\eta_0)=O_p\{n^{-pr/(pr+1)}\}.$$
\end{theorem}

It is shown in Theorem 9.17 of \citet{gu2013smoothing} that the full-basis smoothing spline estimator $\hat{\eta}$ has the convergence rate
$(V+\lambda J)(\hat{\eta}-\eta_0)=O_p\{n^{-pr/(pr+1)}\}$.
Theorem~1 thus states that the proposed estimator $\hat{\eta}_E$ achieves the identical convergence rate as the full-basis estimator.
In particular, the convergence rate of the full-basis estimator gives a lower bound for all the estimators that are based on a subset of the basis functions.
According to \cite{gu2002penalized,ma2015efficient}, and \cite{meng2020more}, all these proposed estimators have the sample convergence rate as the full-basis estimator $\hat{\eta}$, under different conditions.

We emphasize that the goal of Theorem~1 is not to demonstrate that the proposed estimator enjoys a more superior convergence rate.
Instead, Theorem~1 indicates that to achieve such a convergence rate, the proposed estimator requires a relatively smaller $q$, compared with other estimators.
In particular, both the uniform basis selection method \citep{gu2002penalized} and the adaptive basis selection method \citep{ma2015efficient} require $q=O\{n^{2/(pr+1)+\delta}\}$ for an arbitrary small positive number $\delta$.
While for the proposed method, combining the condition $q^{1+2/d}\lambda^{2/r} \rightarrow \infty$ and $\lambda \asymp n^{-r/(pr+1)}$ in Theorem~1 yields, an essential choice of $q$ should satisfy $q=O[n^{2d/\{(pr+1)(d+2)\}+\delta}]$, which is a smaller order of $O\{n^{2/(pr+1)+\delta}\}$.
Although the estimator proposed in \cite{meng2020more} only require $q=O\{n^{(1+\delta)/(pr+1)}\}$, their work assume the sample is uniformly generated from a hypercube, and such an assumption is not always achievable in practice.
In the cases when the sample follows an arbitrary distribution, to the best of our knowledge, the proposed estimator is the one that requires the smallest order of $q$.

Consider the parameter $p$ and $q$ in Theorem~1. It is known that $q$ is associated with the type of the spline, and $p$ is closely associated with $\eta_0$. 
Both parameters have an impact on the convergence rate of the proposed estimator. 
According to \citet{gu2013smoothing}, a common strategy is to set $p\in[1,2]$ and $r\in[4-\delta,4]$ for cubic smoothing splines and tensor-product splines, in which case the size of $q$ roughly lies in the interval $(O(n^{2d/\{9(d+2)\}}), O(n^{2d/\{5(d+2)\}}))$. 
We refer to \citet{gu2013smoothing} for more technical discussion on how to select $p$ and $r$ in practice.

\section{Simulation Results}
\label{section:5}
To show the effectiveness of the proposed smoothing spline estimator, we compare it with three mainstream competitors in terms of prediction accuracy.
The competitors include the uniform basis selection method \citep{gu2002penalized}, the adaptive basis selection method \citep{ma2015efficient,ma2017adaptive}, and the space-filling basis selection method \citep{meng2020more}.
All the methods are implemented in \texttt{R}, and all the parameters are set as default. 

We measure the performance for each method using the prediction mean squared error (MSE), defined as $[\sum_{i=1}^{n}\{\hat{\eta}_E(\bm{t}_i)-\eta_0(\bm{t}_i)\}^2]/n$, where $\{\bm{t}_i\}_{i=1}^{n}$ is an independent testing dataset generate from the same probability density function as the training sample.
Standard errors are calculated through a hundred replicates.
In each replicate, we generate a synthetic training sample with $n=2000$ from each of the following four probability density functions, and the sample is then scaled to $[0,1]^d$,

\begin{itemize}[noitemsep]
    \item \texttt{D1}: Uniform distribution on $[0,1]^d$;
    \item \texttt{D2}: A mixture $t$-distribution $(T_1,\ldots,T_d)$, where $\{T_i\}_{i=1}^d$ are independently generated from $t(10,-5)/2+t(10,5)/2$;
    \item \texttt{D3}: A multivariate Gaussian distribution $\mathcal{N}(0,\Sigma)$, where $\Sigma_{ij}=0.9^{|i-j|}$, $i,j=\{1,\ldots,d\}$;
    \item \texttt{D4}: A banana-shape distribution, which is generated by $(Z_1, Z_2+\frac{Z_1^2}{1.2}, \ldots, Z_d+\frac{Z_1^2}{1.2})$, where $(Z_1,Z_2,\ldots,Z_d)$ is generated from the standard multivariate Gaussian distribution.
\end{itemize}
We consider four different regression functions, which are analogous to the functions considered in \citep{wood2003thin,lin2006component}:
\begin{itemize}[noitemsep]
    \item \texttt{F1}: A 2-d function $\sin(10/(x_1+x_2+0.15))$;
    \item \texttt{F2}: A 2-d function $h_1(x_1,x_2)+h_2(x_1,x_2)$, where $\sigma_1=0.1$, $\sigma_2=0.2$, and
    \begin{eqnarray*}
    h_1(t_1,t_2) = \{0.75/(\pi \sigma_1\sigma_2)\}\times\exp\{-(t_1-0.2)^2/\sigma_1^2 - (t_2-0.3)^2/\sigma_2^2\},\\
    h_2(t_1,t_2) = \{0.75/(\pi \sigma_1\sigma_2)\}\times\exp\{-(t_1-0.7)^2/\sigma_1^2 - (t_2-0.5)^2/\sigma_2^2\};
    \end{eqnarray*}
    \item \texttt{F3}: A 3-d function $\sin(\pi(x_1+x_2+x_3)/3)-x_1-x_2^2$;
    \item \texttt{F4}: A 4-d function
    \begin{align*}
    &x_1+(2x_2-1)^2/2+[\sin(10\pi x_3)/\{2-\sin(10\pi x_3)\}]/3+\\
    &\{0.1\sin(2\pi x_4)+0.2\cos(4\pi x_4)+0.3\sin(6\pi x_4)^2+0.4\cos(8\pi x_4)^3+0.5\sin(10\pi x_4)^3\}/4.
    \end{align*}
\end{itemize}
The signal-to-noise ratio, defined as $\mbox{var}\{\eta(X)\}/\sigma^2$, is set to be two.
We find the results show similar patterns with a large range of signal-noise-ratios.
We set the number of basis $q$ to be $\{20,40,60,80,100\}$. 
To combat the curse-of-dimensionality, we fit smoothing spline analysis of variance models with all main effects and two-way interactions.

\begin{figure}[!ht]\centering
        \includegraphics[scale = 0.9]{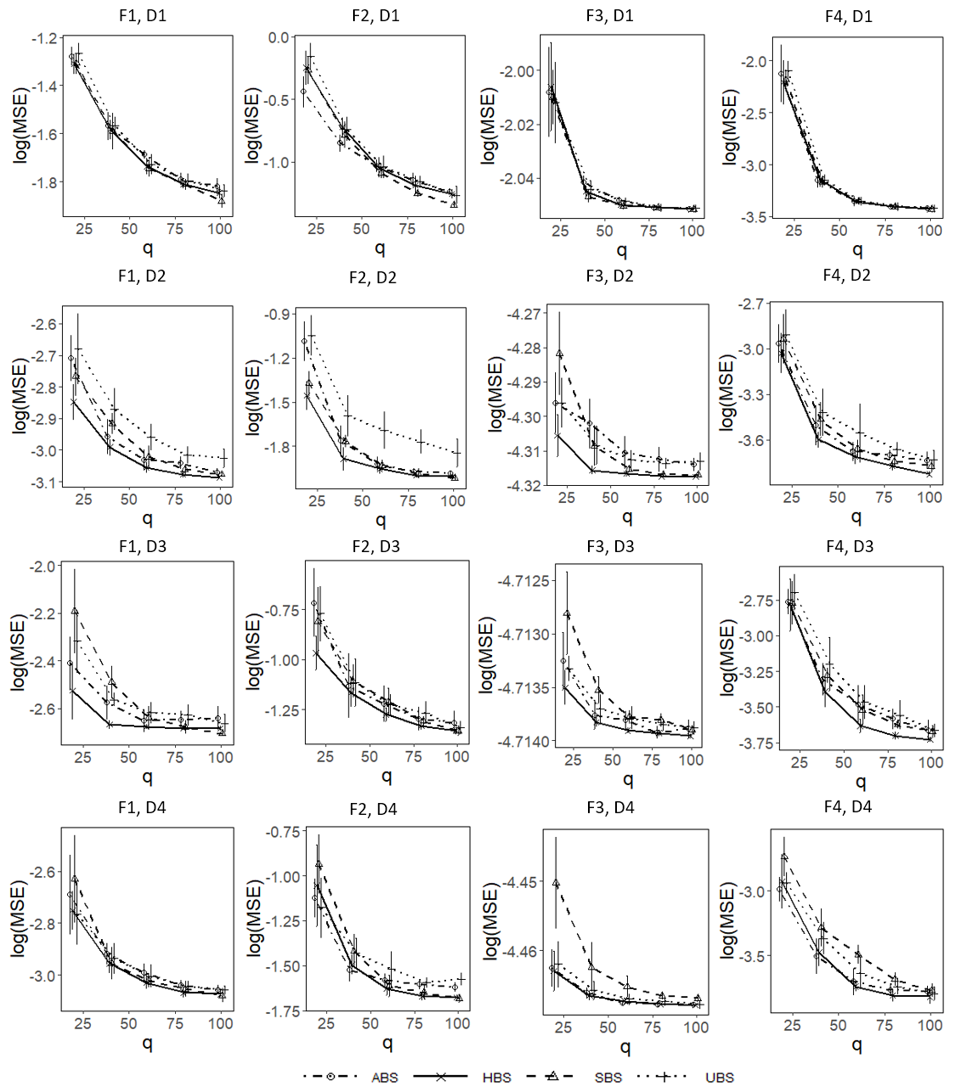} 
        \caption{Simulation under different regression functions (from left to right) and different probability density functions for the predictors (from upper to lower). 
        The prediction errors are plotted versus different $q$.
        Vertical bars represent the standard errors obtained from a hundred replicates.
        } \label{fig4}
\end{figure}

Figure~\ref{fig4} shows the log prediction MSE versus different $q$ under various settings. 
Each row represents a particular data distribution \texttt{D1}--\texttt{D4}, and each column represents a particular regression function \texttt{F1}--\texttt{F4}.
We use solid lines to denote the proposed Hilbert basis selection method (HBS), dash-dotted lines to denote the adaptive basis selection method (ABS), dashed lines to denote the space-filling basis selection method (SBS), and dotted lines to denote the uniform basis selection method (UBS).
The standard error bars are obtained from one hundred replicates.
The results for the full-basis estimator is omitted here due to its high computation cost.

Three significant observations can be made from Fig.~\ref{fig4}.
We first observe that all the methods perform similarly, while the uniform basis selection method performs slightly worse, in the first row of Fig.~\ref{fig4}, in which cases the observations are uniformly-distributed in a hypercube.
Such an observation is consistent with the simulation results in \citet{meng2020more}, which suggests both the space-filling basis selection method and the adaptive basis selection method consistently outperform the uniform basis selection method.
Nevertheless, in the lower three rows of Fig.~\ref{fig4}, we observe the uniform basis selection method yields decent performance occasionally.
Such an observation suggests when the predictors do not follow the uniform distribution on a hypercube, none of the four basis selection methods consistently dominates the others.

Second, the MSE for the proposed estimator decreases faster than the other estimators as $q$ increase.  
This observation is consistent with Theorem 1, which suggests the proposed estimator requires smaller $q$ to achieve the identical convergence rate as the full-basis estimator.

Third, the proposed estimator may suffer from deteriorating performance when $q$ is too small. 
We attributed such an observation to the fact that, when $q$ is small, the proposed method tends to select a large proportion of basis functions corresponding to the data points that are close to the boundary. 
These basis functions may not have adequate benefits in terms of prediction.
As $q$ increases, the proportion of such basis functions decreases, and thus the proposed estimator achieves better performance.
It is suggested in \cite{gu2014smoothing} to let $q\geq30$ for robust prediction in practice.
According to such a suggestion and consider the cases when $q\geq30$ in Fig~\ref{fig4}, we observe the proposed estimator outperforms the competitors in most of the settings.

\section{Real data example}
\label{section:6}
\begin{figure}[!ht]
\centering{
\includegraphics[scale = .7]{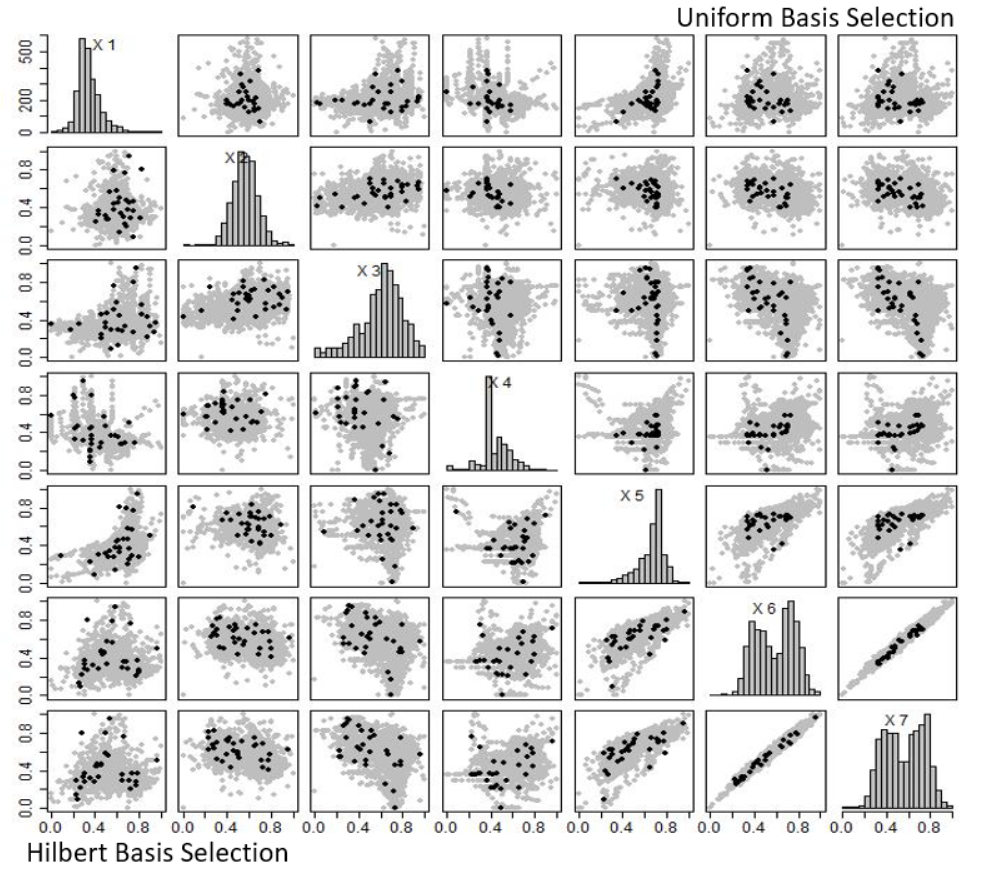}
}
\caption{The diagonal panels show histograms for each of the predictors. The off-diagonal panels show the scatter plots corresponding to each pair of the predictors. The bases selected by the proposed method and the uniform basis selection method are shown in the lower diagonal panels and the upper diagonal panels, respectively. The black dots are the observations corresponding to the selected basis functions when $q=40$. }
\label{fig5}
\end{figure}

In petroleum refinery, a debutanizer column is used to separate butane from gasoline. Estimating the butane concentration in the bottom product of the debutanizer column is essential for improving the performance of the refining process. 
Of interest is to predict the butane concentration using the conditions of debutanizer columns and other related information, which are measured by the soft sensors in the petroleum refinery process. 
We consider a dataset with $n = 2,395$ and seven predictor variables, includes temperature, pressure, flow, and so on \footnote{The dataset can be downloaded from https://home.isr.uc.pt/~fasouza/datasets.html}. 
More details of this dataset can be found in \cite{fortuna2007soft}.
The sample is first scaled to $[0,1]^7$ as a preprocessing step.
Figure~\ref{fig5} shows histograms for each of the predictors in diagonal panels and scatter plots for each pair of the predictors in off-diagonal panels. 
We observe the sample in this dataset is extremely non-uniformly-distributed.
The basis functions selected by the proposed Hilbert basis selection method and the uniform basis selection method are marked as black dots in the lower diagonal panels and the upper panels, respectively.
Compare with the uniform basis selection method, we observe the proposed method selects the basis functions corresponding to the observations that are more equally-spaced.

\begin{figure}[ht]
\centering{
    \includegraphics[scale = 1]{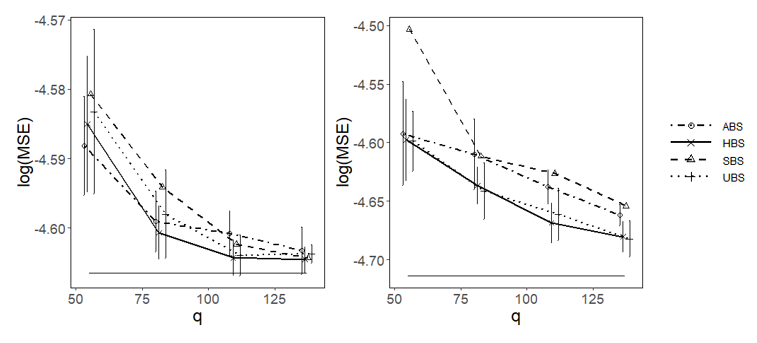}
   }
\caption{The left panel shows the log prediction MSE versus different $q$ under model setting M1 for the debutanizer column dataset. The right panel shows the log prediction MSE versus different $q$ under model setting M2. The horizontal lines represent the performance for the full sample estimator.}
\label{fig6}
\end{figure}

We fitted the cubic tensor product smoothing spline analysis of variance model to the dataset, and we considered two different model settings,
\begin{itemize}
\item \texttt{M1}: additive model,
$$y_i= \eta_\varnothing +\sum_{j=1}^7\eta_j(x_{ij})+\epsilon_i, \quad i=1,\ldots,n;$$
\item \texttt{M2}: by the preliminary model diagnostics \citep{gu2004model}, we considered the following functional ANOVA decomposition,
\begin{equation*}
\begin{aligned}
y_i= & \eta_\varnothing +\sum_{j=1}^7\eta_j(x_{ij}) + \eta_{1,3}(x_{i1},x_{i3}) 
+ \eta_{1,5}(x_{i1},x_{i5}) + \eta_{1,6}(x_{i1},x_{i6}) \\
&+ \eta_{3,5}(x_{i3},x_{i5})  +\epsilon_i, \quad i=1,\ldots,n. 
\end{aligned}
\end{equation*}
\end{itemize}
Here, the response $y_i$ is the butane concentration of $i$-th observation, $x_{ij}$ is the value of the $j$-th predictor of the $i$-th observation, $\eta_\varnothing$ is a constant function, $\{\eta_j\}_{j=1}^7$ are main effect functions, $\eta_{1,3}, \eta_{1,5}, \eta_{1,6}, \eta_{3,5}$ are two-way interaction functions of corresponding predictors, and $\epsilon_i$'s are i.i.d. normal errors with zero mean and unknown variance. 
We replicated the experiment one hundred times. 
To show the effectiveness of the proposed estimator, we compared it with the other three mainstream competitors, as mentioned in the previous section, in terms of the prediction MSE calculated on a holdout testing set.
Figures~\ref{fig6} shows the log prediction MSE versus different $q$ under two different model settings. 
Vertical bars represent the standard errors obtained from a hundred replicates.
The horizontal lines represent the performance for the full sample estimator.
We observe that the proposed estimator, labeled as solid lines, yields the second-best result for the smallest $q$ considered here and the best result for other cases. 
We attribute such an observation to the fact that the proposed method selects the basis functions corresponding to roughly equally-spaced observations, resulting in a more effective estimation of the underlying regression function.

\section{Discussion}
\label{section:7}

In this paper, we proposed a novel basis selection method for smoothing splines approximation.
Unlike the existing basis selection approaches, which mainly focus on the setting that the sample is uniformly distributed on a unit hypercube, the proposed method aims to provide an effective estimation that is adaptive to an arbitrary probability distribution of the sample.
Motivated by importance sampling, we achieved the goal by carefully selecting a set of approximately equally-spaced observations, even when the sample is not uniformly-distributed.
We proposed an efficient algorithm for identifying such observations by utilizing the Hilbert space-filling curve.
The proposed estimator has the same convergence rate as the full-basis estimator when the number of basis $q$ is roughly at the order of $O[n^{2d/\{(pr+1)(d+2)\}}]$.
The superior performance of HBS over mainstream competitors was justified by various numerical experiments.

Our work is related to \citet{meng2020more}.
In particular, \citet{meng2020more} utilized space-filling design techniques, or low-discrepancy sequences, to identify an approximately equally-spaced subsample from ``uniformly-distributed" sample, resulting in an efficient approximation.
Our work extended their work to the non-uniform distribution setting and theoretically showed that the basis functions corresponding to an equally-spaced subsample still benefits the smoothing splines approximation.

The proposed method has the penitential to be applied to many large-sample applications, including but not limited to Gaussian process regression, kernel ridge regression, and low-rank approximation of matrices.
This work may speed up these techniques with theoretical guarantees.
Some additional simulation results are provided in Supplementary Material.

\section*{Acknowledgment}
The authors thank the associate editor and two anonymous reviewers for provided helpful comments on earlier drafts of the manuscript.
The authors would like to acknowledge the support from the U.S. National Science Foundation under grants DMS-1903226, DMS-1925066, the U.S. National Institute of Health under grant R01GM122080, National Natural Science Foundation of China Grant No.12101606, No.12001042, and Beijing Institute of Technology research fund program for young scholars.


\newpage
\bibliographystyle{plainnat}
\bibliography{ref}
\end{document}


\title{Supplementary materials for "Smoothing splines approximation using Hilbert curve basis selection"}
\date{}

\author[1]{Cheng Meng}
\author[2]{Jun Yu}
\author[3]{Yongkai Chen}
\author[3]{Wenxuan Zhong}
\author[3,*]{Ping Ma}

\affil[1]{Institute of Statistics and Big Data, Renmin University of China}
\affil[2]{School of Mathematics and Statistics, Beijing Institute of Technology}
\affil[3]{Department of Statistics, University of Georgia}
\affil[*]{Corresponding author: Ping Ma, pingma@uga.edu}

\maketitle

\begin{description}

\item[Title:] Proofs of theoretical results and related materials.

\item[Formal definition of Hilbert curves:] We present formal definition of Hilbert curves with details.

\item[Proof of Lemma 4.1:] A proof of Lemma 4.1.

\item[Proof of essential lemmas:] We provide the proof of some lemmas that are essential to Theorem~1.

\item[Proof of Theorem 4.1:] A proof of the main theorem.

\item[Additional simulation results:] We provide additional simulation results to show the effect of the $k$ in Algorithm~1. In addition, we show some results that indicating the proposed method could also be utilized to accelerate the estimation of kernel ridge regression.

\item[Dataset:] Data are publicly available.

https://home.isr.uc.pt/ fasouza/datasets.html

\end{description}

\appendix

\section{Formal definition of Hilbert curves}

Recall that $d$ is the sample dimension. 
For a positive integer $k$, we define $2^{dk}$ intervals $I_d^k(i)=[i/2^{dk},(i+1)/2^{dk}]$, $i=0,\ldots,2^{dk}-1$, and let $\mathcal{I}_d^k=\{I_d^k(i)|0\leq i<2^{dk} \}$.
For $\kappa=(i_1,\ldots,i_d)$ with $i_j\in\{0,\ldots,2^k-1\}$, we then define $2^{dk}$ subcubes of $[0,1]^d$ via $E_d^k(\kappa)=\prod_{j=1}^d[i_j/2^k, (i_j+1)/2^k]$, and let $\mathcal{E}_d^k=\{E_d^k(\kappa)|\kappa\in\{0,\ldots,2^k-1\}^d\}$.
The $k$th order $d$-dimensional Hilbert space-filling curve $H_k:\mathcal{I}_d^k\rightarrow\mathcal{E}_d^k$ is defined as a map that connects the regular grid $\mathcal{I}_d^k$ to the subcubes $\mathcal{E}_d^k$ in some certain order.
A general geometric procedure that allows the construction of an entire class of the $H_k$ can be found in \cite{zumbusch2012parallel}.
It can be shown that $H_k:\mathcal{I}_d^k\rightarrow\mathcal{E}_d^k$ is a well-defined bijective mapping.
Informally, we call $\{H_1,H_2,\ldots\}$ the Hilbert space-filling curves.
The Hilbert curve is defined by $H(x)=\lim_{k\rightarrow\infty}H_k(x)$, which is a mapping from $[0,1]$ to $[0,1]^d$.
Several important properties of the Hilbert curve $H$ are presented below, see \cite{zumbusch2012parallel} for further details.
\begin{enumerate}[noitemsep]
    \item $H[I_d^k(i)]=H_k[I_d^k(i)]$.
    \item Let $\lambda_d$ be the $d$-dimensional Lebesgue measure. For any measurable set $\Omega\subseteq[0,1]$, one has $\lambda_1(\Omega)=\lambda_d(H(\Omega))$.
    \item If a random variable $X$ follows the uniform distribution on $[0,1]$, then $H(X)$ follows the uniform distribution on $[0,1]^d$. This is to say, we have $\int_{[0,1]^d}f(\bm{x})\mbox{d}\bm{x}=\int_0^1f[H(x)]\mbox{d}x$.
    \item For any $x,y\in[0,1]$, we have $||H(x)-H(y)||\leq2\sqrt{d+3}|x-y|^{1/d}$. 
\end{enumerate}

\section{Proof of Lemma 4.1}
\subsection{Proof of the asymptotically unbiased estimate}\label{sec:b1}

Recall that $\{\x_i^*\}_{i=1}^q$ represents the selected subsample from the sample $\{\x_i\}_{i=1}^n$.
We start by showing that $\sum_{j=1}^q\{\tilde{n}_j/n\}\phi_\nu(\bm{x}_j^*)\phi_\mu(\bm{x}_j^*)$ is an asymptotically unbiased estimate for $\int_{\Omega}\phi_\nu(\bm{x})\phi_\mu(\bm{x}) f_X(\bm{x})\mbox{d}\bm{x}$.
Let the partition $\cup_i I_i, i=1,\ldots,C$ be the bins of the histogram in Step~3 of Algorithm~1.
Without lose of generality, we assume $C=2^{dk}, E_i=H_k(I_d^k(i)), i=1,\ldots, 2^{dk}$.
Let $I_{E_i}(\x)$ be the indicator function, such that, if $\x$ lies in $E_i$, it equals one; on the contrary, it equals zero.
Let $n_i$ be the number of observations in $E_i$, $i=1,\ldots, 2^{dk}$.
Recall that a simple random sampling procedure is conducted in strata $E_i$ in Algorithm~1, and $E_i$ does not depend on the observed sample.
As a result, for any $i\in\{1,\ldots, 2^{dk}\}$, we have 
\begin{eqnarray}\label{SS0}
    E\left[\sum_{j=1}^q\phi_\nu(\x_j^*)\phi_\mu(\x_j^*)I_{E_i}(\x_j^*)\Bigg| \{\x_i\}_{i=1}^n\right]= \frac{1}{n_i}\sum_{j=1}^n\phi_\nu(\x_j)\phi_\mu(\x_j)I_{E_i}(\x_j).
\end{eqnarray}
Through some simple algebra, Equation~(\ref{SS0}) indicates
\begin{eqnarray}\label{SS1}
E\left[\sum_{i=1}^C\sum_{j=1}^q\frac{n_i}{n} \phi_\nu(\x_j^*)\phi_\mu(\x_j^*)I_{E_i}(\x_j^*)\Bigg| \{\x_i\}_{i=1}^n\right]=\frac{1}{n}\sum_{j=1}^n\phi_\nu(\x_j)\phi_\mu(\x_j).
\end{eqnarray}
Recall that $\tilde{n}_j$ is the number of data points within the bin that $\x_j^*$ lies in and only one data point is randomly selected from each non-empty bin. 
Therefore, it is clear that 
\begin{equation*}
\sum_{i=1}^C (n_i/n)\phi_\nu(\bm{x}_j^*)\phi_\mu(\bm{x}_j^*)I_{E_i}(\x_j^*)=
\begin{cases}
      ({\tilde{n}_j}/{n})\phi_\nu(\bm{x}_j^*)\phi_\mu(\bm{x}_j^*) & \text{if $\bm{x}_j^*\in E_i$};\\
      0 & \text{if $ \bm{x}_j^*\not\in E_i$}.\\
\end{cases}       
\end{equation*}
Consequently, we have
\begin{equation}\label{eq:adds2}
  \sum_{i=1}^C\sum_{j=1}^q\frac{n_i}{n} \phi_\nu(\bm{x}_j^*)\phi_\mu(\bm{x}_j^*)I_{E_i}(\x_j^*)=\sum_{j=1}^q\sum_{i=1}^C\frac{n_i}{n} \phi_\nu(\bm{x}_j^*)\phi_\mu(\bm{x}_j^*)I_{E_i}(\x_j^*)=\sum_{j=1}^q\frac{\tilde{n}_j}{n}\phi_\nu(\bm{x}_j^*)\phi_\mu(\bm{x}_j^*).  
\end{equation}
Combining Equation~(\ref{SS1}) and Equation~(\ref{eq:adds2}), we have
\begin{equation}\label{eq:adds3}
E\left[\sum_{j=1}^q({\tilde{n}_j}/{n})\phi_\nu(\bm{x}_j^*)\phi_\mu(\bm{x}_j^*)\right] = E\left[\frac{1}{n}\sum_{j=1}^n\phi_\nu(\x_j)\phi_\mu(\x_j)\right].
\end{equation}
As a result, $\sum_{j=1}^q({\tilde{n}_j}/{n})\phi_\nu(\bm{x}_j^*)\phi_\mu(\bm{x}_j^*)$ is an asymptotically unbiased estimate for $\int_{\Omega}\phi_\nu(\bm{x})\phi_\mu(\bm{x}) f_X(\bm{x})\mbox{d}\bm{x}$.
We now derive the convergence rate for $\sum_{j=1}^q({\tilde{n}_j}/{n})\phi_\nu(\bm{x}_j^*)\phi_\mu(\bm{x}_j^*)$ in the following.

\subsection{Proof of the convergence rate}
For $i=1,\ldots, 2^{dk}$, let $|I_i|$ be the length of the interval $I_i$.
Note that $C$ represents the number of partitions of $[0,1]$, and thus we have $|I_i|=1/C$.
Recall the four properties of the Hilbert space-filling curves in Section~A.
Utilizing the fact that $H_k^{-1}(E_i)=I_i$, along with the first and the fourth property, for any $\bm{x},\bm{y}\in E_i$, $i=1,\ldots, 2^{dk}$, we have
\begin{equation*}
\|\bm x-\bm y\|\le 2(d+3)^{-1/2}|I_i|^{1/d}=2(d+3)^{-1/2}C^{-1/d}.
\end{equation*}
Consequently, under Condition 4, for any $\bm x,\bm y\in E_i$, $i=1,\ldots, 2^{dk}$, there exists a positive constant $M$ such that
\begin{equation}\label{eq:ss1}
|\phi_\nu(\x)\phi_\mu(\x)-\phi_\nu(\y)\phi_\mu(\y)|\le M\|\x-\y\|\le 2M(d+3)^{-1/2}C^{-1/d}.
\end{equation}
Recall that $n_i$ represents the number of data points located in $E_i$.
Inequality~\eqref{eq:ss1} indicates, for any $\bm{x}_j^* \in E_i$, $i=1,\ldots, 2^{dk}$, we have 
\begin{eqnarray*}
&~&\left|\phi_\nu(\x_j^*)\phi_\mu(\x_j^*)I_{E_i}(\x_j^*)-\frac{1}{n_i}\sum_{j=1}^n\phi_\nu(\bm{x}_j)\phi_\mu(\bm{x}_j)I_{E_i}(\x_j)\right|\\
&=&\left|\frac{1}{n_i}n_i\phi_\nu(\x_j^*)\phi_\mu(\x_j^*)I_{E_i}(\x_j^*)-\frac{1}{n_i}\sum_{j=1}^n\phi_\nu(\bm{x}_j)\phi_\mu(\bm{x}_j)I_{E_i}(\x_j)\right|\\
&\le& \frac{1}{n_i}\sum_{\x_j^*,\x_j \in E_i}\left|\phi_\nu(\x_j^*)\phi_\mu(\x_j^*)I_{E_i}(\x_j^*)-\phi_\nu(\bm{x}_j)\phi_\mu(\bm{x}_j)I_{E_i}(\x_j) \right|\\
&\le& 2M(d+3)^{-1/2}C^{-1/d}.    
\end{eqnarray*}
That is to say, the conditional variance of $({\tilde{n}_j}/{n})\phi_\nu(\bm{x}_j^*)\phi_\mu(\bm{x}_j^*)I_{E_j}(\x_j^*)$ can be bounded by
\begin{align}\label{SSS2}
{\rm var}\left[\frac{\tilde{n}_j}{n}\phi_\nu(\bm{x}_j^*)\phi_\mu(\bm{x}_j^*)I_{E_j}(\x_j^*) \Bigg| \{\bm{x}_i\}_{i=1}^n\right] & = \left(\frac{\tilde{n}_j}{n}\right)^2\left[\phi_\nu(\bm{x}_j^*)\phi_\mu(\bm{x}_j^*)I_{E_i}(\x_j^*)-\frac{1}{n_i}\sum_{j=1}^n\phi_\nu(\bm{x}_j)\phi_\mu(\bm{x}_j)I_{E_i}(\x_j)\right]^2\nonumber\\
& \le \left(\frac{\tilde{n}_j}{n}\right)^2 4M^2(d+3)^{-1}C^{-2/d}.    
\end{align}
Consequently, the conditional variance of $\sum_{j=1}^C({\tilde{n}_j}/{n})\phi_\nu(\bm{x}_j^*)\phi_\mu(\bm{x}_j^*)I_{E_j}(\x_j^*)$ can be represented by
\begin{align}\label{eq:s3}
{\rm var}\left[\sum_{j=1}^C\frac{\tilde{n}_j}{n}\phi_\nu(\bm{x}_j^*)\phi_\mu(\bm{x}_j^*)I_{E_j}(\x_j^*) \Bigg| \{\bm{x}_i\}_{i=1}^n\right]   = & {\rm var}\left[\sum_{j=1}^q\frac{\tilde{n}_j}{n}\phi_\nu(\bm{x}_j^*)\phi_\mu(\bm{x}_j^*) \Bigg| \{\bm{x}_i\}_{i=1}^n\right]\nonumber\\
     \nonumber\le&\sum_{j=1}^q \left(\frac{\tilde{n}_j}{n}\right)^2 4M^2(d+3)^{-1}C^{-2/d}\\
     \nonumber\le& \max_{i\in\{1,\ldots,C\}}\left\{\left(\frac{n_i}{n}\right)^2\right\}\times q\times 4M^2(d+3)^{-1}C^{-2/d}\\
     \nonumber=& \Big(O_p(1)/q\Big)\times4M^2(d+3)^{-1}C^{-2/d}\\
    =&O_p(q^{-1-2/d}),
\end{align}
where the first equality holds by the fact that we do not take any sample on the empty strata, the first inequality follows from Inequality~(\ref{SSS2}), the second last equality follows from Condition~5, and the last equality follows from the scenario of our interest that $q\le C$.
Recall that under Condition~4, we have
\begin{equation}\label{eq:s4}
\left|\frac{1}{n}\sum_{j=1}^n\phi_\nu(\bm{x}_j)\phi_\mu(\bm{x}_j)-\int_{\Omega}\phi_\nu(\bm{x})\phi_\mu(\bm{x}) f_X(\bm{x}) \mbox{d}\bm{x}\right|^2=O_p(n^{-1}).
\end{equation}
Combining Equation~(\ref{eq:adds3}), Inequality (\ref{eq:s3}), and Equation~(\ref{eq:s4}), we have
\begin{align}\label{eq:s5}
\nonumber &{\rm var}\left\lbrack\sum_{j=1}^C\frac{\tilde{n}_j}{n}\phi_\nu(\bm{x}_j^*)\phi_\mu(\bm{x}_j^*)\right\rbrack\\
\nonumber =&E\left\{{\rm var}\left\lbrack\sum_{j=1}^C\frac{\tilde{n}_j}{n}\phi_\nu(\bm{x}_j^*)\phi_\mu(\bm{x}_j^*) \Bigg| \{\bm{x}_i\}_{i=1}^n\right\rbrack\right\}+{\rm var}\left\{ E\left[\sum_{j=1}^C\frac{\tilde{n}_j}{n}\phi_\nu(\bm{x}_j^*)\phi_\mu(\bm{x}_j^*)\Bigg| \{\bm{x}_i\}_{i=1}^n\right]\right\}\\
\nonumber    =&O_p(q^{-1-2/d})+O_p(n^{-1})\\
    =&O_p(q^{-1-2/d}),
\end{align}
where the last equation follows from Condition 6.
Together with Equation~(\ref{eq:s4}) and Equation~(\ref{eq:s5}), the desired result follows by the Holder's inequality, i.e.,
\begin{align*}
&\left\lbrack\int_{\Omega}\phi_\nu(\bm{x})\phi_\mu(\bm{x}) f_X(\bm{x}) \mbox{d}\bm{x}-\sum_{j=1}^q\frac{\tilde{n}_j}{n}\phi_\nu(\bm{x}_j^*)\phi_\mu(\bm{x}_j^*)\right\rbrack^2\\
\le& 2\left\lbrack\int_{\Omega}\phi_\nu(\bm{x})\phi_\mu(\bm{x}) f_X(\bm{x}) \mbox{d}\bm{x}-\frac{1}{n}\sum_{j=1}^n\phi_\nu(\bm{x}_j)\phi_\mu(\bm{x}_j)\right\rbrack^2\\
&+2\left\lbrack\frac{1}{n}\sum_{j=1}^n\phi_\nu(\bm{x}_j)\phi_\mu(\bm{x}_j)-\sum_{j=1}^q\frac{\tilde{n}_j}{n}\phi_\nu(\bm{x}_j^*)\phi_\mu(\bm{x}_j^*)\right\rbrack^2\\
=&O_p(n^{-1})+O_p(q^{-1-2/d})\\
=&O_p(q^{-1-2/d}).
\end{align*}

\section{Proof of essential lemmas}
Recall that for basis selection method, we approximate the smoothing spline estimator in the effective model space $\H_E$. 
Let $\H\ominus \H_E$ be the orthogonal complement of $\H_E$ in the reproducing kernel Hilbert space $\H$. 
We have the following lemma which justifies the use of the effective space $\H_E$.
\begin{lemma}\label{lem_s1}
Under Conditions $1-5$, as $\lambda\rightarrow 0$ and $q^{1+2/d}\lambda^{2/r} \rightarrow \infty$, $\forall h\in \mathcal{H} \ominus \mathcal{H}_E$, we have $V(h)=o_p\{\lambda J(h)\}$.
\end{lemma}
Before we prove the lemma, we first introduce an essential lemma as follows, refer to Lemma 9.1 of \citet{gu2013smoothing} for details.
\begin{lemma}\label{lem_s2}
Under Condition 2, as $\lambda\rightarrow 0$, one has
\begin{align*}
&\sum_\nu\frac{\lambda\rho_\nu}{(1+\lambda\rho_\nu)^2}=O(\lambda^{-1/r}),\\
&\sum_\nu\frac{1}{(1+\lambda\rho_\nu)^2}=O(\lambda^{-1/r}),\\
&\sum_\nu\frac{1}{1+\lambda\rho_\nu}=O(\lambda^{-1/r}).
\end{align*}
\end{lemma}

\begin{proof}
For $h\in\H\ominus \H_E$, one has $h(\bm{x}_i)=J(R_J(\bm{x}_i,\cdot),h)=0$ for any $i\in\{1,\ldots,n\}$.
Consequently, we have $\sum_{j=1}^q ({\tilde{n}_j}/{n})h^2(\bm{x}_j^*)=0$. 
Write $h=\sum_\nu h_\nu\phi_\nu$, it follows that
\begin{align}
V(h)&=\int_{\Omega}h(\bm{x})^2 f_X(\bm{x})\mbox{d}\bm{x} \nonumber\\
&=\sum_\nu\sum_\mu h_\nu h_\mu \int_{\Omega}\phi_\nu(\bm{x})\phi_\mu(\bm{x}) f_X(\bm{x}) \mbox{d}\bm{x} \nonumber\\
&=\sum_\nu\sum_\mu h_\nu h_\mu \left\lbrack\int_{\Omega}\phi_\nu(\bm{x})\phi_\mu(\bm{x}) f_X(\bm{x}) \mbox{d}\bm{x}-\sum_{j=1}^q\frac{\tilde{n}_j}{n}\phi_\nu(\bm{x}_j^*)\phi_\mu(\bm{x}_j^*)\right\rbrack.\label{S18}
\end{align}

The Cauchy inequality yields.
\begin{align}
\mbox{(\ref{S18})}&\leq \left\lbrack\sum_\nu\sum_\mu \frac{1}{1+\lambda\rho_\nu}\frac{1}{1+\lambda\rho_\mu}
\left(\int_{\Omega}\phi_\nu(x)\phi_\mu(\bm{x}) f_X(\bm{x}) \mbox{d}\bm{x}-\sum_{j=1}^q\frac{\tilde{n}_j}{n}\phi_\nu(\bm{x}_j^*)\phi_\mu(\bm{x}_j^*)\right)^2
\right\rbrack^{1/2} \nonumber\\
&\times \left( \sum_\nu\sum_\mu (1+\lambda\rho_\nu)(1+\lambda\rho_\mu)h_\nu^2 h_\mu^2\right)^{1/2}.\label{eqn30}
\end{align}

By Lemma \ref{lem_s2}, one has
\begin{eqnarray}\label{eqn56}
\sum_\nu\sum_\mu \frac{1}{1+\lambda\rho_\nu}\frac{1}{1+\lambda\rho_\mu} = O(\lambda^{-2/r}),
\end{eqnarray}
and Lemma 4.1 shows
\begin{eqnarray}\label{eqn28}
\left( \int_{\Omega}\phi_\nu(\bm{z})\phi_\mu(\bm{z}) f_X(\bm{z}) \mbox{d}\bm{z}-\sum_{j=1}^q\frac{\tilde{n}_j}{n}\phi_\nu(\bm{z}_j^*)\phi_\mu(\bm{z}_j^*)\right)^2 = O_p(q^{-1-2/d}).
\end{eqnarray}
One also has 
\begin{eqnarray}\label{eqn29}
\sum_\nu (1+\lambda\rho_\nu)h_\nu^2 = (V+\lambda J)(h),
\end{eqnarray}
since $\phi_\nu$'s simultaneously diagonalize $V$ and $J$.
Combining the results in (\ref{eqn30}), (\ref{eqn56}), (\ref{eqn28}), and (\ref{eqn29}), we have
\begin{eqnarray}\label{eqn34}
V(h)\le \{O_p(q^{-1-2/d})\lambda^{-2/r}\}(V+\lambda J)(h).
\end{eqnarray}
As a result, when $q^{1+2/d}\lambda^{2/r} \rightarrow \infty$, Inequality (\ref{eqn34}) yields
\begin{eqnarray*}
V(h)=o_p\{\lambda J(h)\}.
\end{eqnarray*}
\end{proof}

\section{Proof of Theorem 4.1}
Compared with the condition from $q\lambda^{2/r} \to \infty$ in Theorem 9.17 in \cite{gu2013smoothing}, 
the condition for Theorem 1 is $q^{1+2/d}\lambda^{2/r} \rightarrow \infty$. As a result, the proposed smoothing spline estimator $\hat{\eta}_E$ can have the same convergence rate as the full-basis estimator with smaller $q$. Under this condition as well as Conditions 1-5, as $\lambda \to 0$, $\forall h\in \mathcal{H} \ominus \mathcal{H}_E$, $V(h)$ is dominated by $\lambda J(h)$, which is guaranteed by Lemma~\ref{lem_s2}. 
Theorem 1 thus can be proved by following the proof of Theorem 9.17 in \cite{gu2013smoothing} directly.

\section{Additional simulation results}

\begin{figure}[ht]
\centering{
    \includegraphics[scale = 0.8]{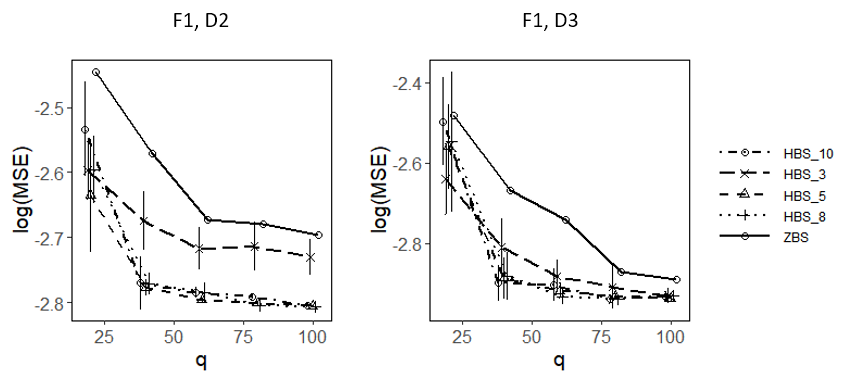}
   }
\caption{Simulation under the setting ``F1, D2" and ``F1, D3".}
\label{fig7}
\end{figure}

We provide more simulation results to show the effect of different choices of space-filling curves and $k$ in Algorithm~1.
The simulation settings are the same as the ones in Section~5.
We considered four different choices of Hilbert space-filling curves w.r.t. $k=3,5,8,10$. 
We also considered the Z-order curve, which is another popular representative of space-filling curves.
The Z-order curve is denoted by ZBS.
Figure~\ref{fig7} shows the result w.r.t. the simulation setting ``F1, D2" and ``F1, D3".
The results w.r.t. other settings show similar patterns, thus are omitted here.
Three significant observations can be made from Fig.~\ref{fig7}.
First, ZBS performs uniformly worse than HBS methods.
Such an observation is expected, as noted in \cite{he2016extensible} that, compared to Hilbert curves, other space-filling curves attain the same asymptotic rates, whereas with a worse constant.
Second, we observe that HBS\_3 performs worse than other HBS methods.
This is because $k=3$ is too small for such cases.
Finally, we observe that the performance of HBS\_5, HBS\_8, and HBS\_10 are almost the same.
Such an observation is also expected, as we noted in Section~3 that as long as the number of bins is fixed, increasing the value of $k$ does not affect the result.

\begin{figure}[ht]
\centering{
    \includegraphics[scale = 1]{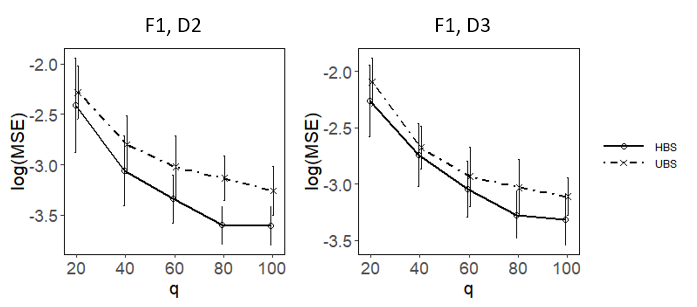}
   }
\caption{Simulation under the setting ``F1, D1" and ``F1, D2" using the kernel ridge regression.}
\label{fig8}
\end{figure}

Although we mainly focus on smoothing splines in this paper, it is possible that the proposed method could also accelerate the estimation of other nonparametric regression estimators, includes the thin plate regression splines \citep{wood2003thin} and the kernel ridge regression \citep{yang2017randomized}, and etc.
To support this claim, we apply the proposed basis selection approach to accelerate the computational of kernel ridge regression, following the procedure of \cite{yang2017randomized}.
Figure~\ref{fig8} shows the results w.r.t. the simulation setting ``F1, D2" and ``F1, D3".
The results w.r.t. other settings show similar patterns, thus are omitted here.
These results indicate the basis selection method has the potential to accelerate the approximation of various nonparametric regression approaches.
Furthermore, compared to the uniform basis selection method, the proposed Hilbert basis selection method tends to be more effective.

\bibliographystyle{plainnat}
\bibliography{ref}